\newcommand{\doublespace}{
  \renewcommand{\baselinestretch}{1.5}
  \large\normalsize}
\newcommand{\singlespace}{
  \renewcommand{\baselinestretch}{1.25}
  \large\normalsize}
\documentstyle[preprint,aps,tighten,prb]{revtex}
\begin{document}
\doublespace
\title{Optical properties 
of self-assembled quantum wires for application in infra-red detection}

\author{Liang-Xin Li, Sophia Sun, and Yia-Chung Chang}
\address{Department of Physics and Materials Research Laboratory 
\\ University of Illinois at Urbana-Champaign, Urbana, Illinois 61801}

\date{\today}
\maketitle

\begin{abstract}

We present theoretical studies of optical properties of
Ga$_{1-x}$In$_x$As self-assembled quantum-wires (QWR's) made of 
short-period superlattices with strain-induced lateral ordering. 
Valence-band anisotropy, band mixing, and effects due to 
local strain distribution at the atomistic level are all taken into account.
Using realistic material parameters which are experimentally feasible, we
perform simulations of the absorption spectra for both inter-subband
and inter-band transitions (including the excitonic effect) of this material. 
It is shown that the self-assembled QWR's have favorable optical properties
for application in infra-red detection with normal incidence.
The wavelength of detection ranges from
10 $\mu m$ to 20 $\mu m$ with the length of QWR period varying from 
150\AA to 300\AA.


\end{abstract}

\mbox{}

\mbox{}

\newpage

\section{Introduction}

	Quantum-well infra-red photodetectors (QWIP's) have been
extensively studied in recent years. The main mechamism used in
QWIPs is the inter-subband optical transition, because the wavelengths
for these transitions in typical III-V quantum wells can be tailored
to match the desired operating wavelength (1-20 $\mu m$)
for infra-red (IR) detection. Due to its narrow band absorption,
QWIP's are complementary to the traditional HgCdTe detectors, which
utilize the inter-band absorption, and therfeore are applicable 
only for broad-band absorption.
The main drawback of QWIP's is the lack of normal-incidence
capability, unless some processing is made to create 
diffraction gartings on the surface, which tends to reduce the responsivity
of the material to the incident radiation. Because electrons in
quantum wells have translational invariance (within the effective-mass model)
in the plane normal to the growth axis, the electron
inter-subband transitions for normal-incident radiation is zero 
(or very small even if the coupling with other bands is considered).
One way to break the translational invariance is to introduce the
surface diffraction grating as commonly adopted in many QWIP's fabricated today.
A better (and less expensive) way to break the in-plane translational
invariance is to utilize the strain-induced lateral modulation provided in 
self-assembled nano-structure materials. These nano-structures inculde
quantum dots and quantum wires. Because the lateral modelation is formed
via self-assembly, the fabrication of this type of materials will be much 
more efficient once the optimized growth parameters are known. Hence, it will be
cost effective to use them for device fabrications.

Self-assembled III-V QWR's via the strain-induced lateral-layer ordering
(SILO) process have attracted a great deal of attention recently.
[\cite{chou}$^-$\cite{wohlert1}] 
The self-assembly process occurs during the growth of
short-period superlattices (SPS) [e.g. (GaAs)$_2$/(InAs)$_{2.25}$]
along the [001] direction on InP substrate. 
The excess fractional InAs layer leads to stripe-like islands 
during the initial MBE growth.[4] The presence of stripes combined with strain
leads to natural phase separation as
additional layers of GaAs or InAs are deposited and the structure becomes
laterally modulated in terms of In/Ga composition.
A self-assembled QWR heterostructure can then be
created by sandwiching the laterally modulated layer
between barrier materials such as Al$_{0.24}$Ga$_{0.24}$In$_{0.52}$As 
(quarternary), Al$_{0.48}$In$_{0.52}$As (ternary), or InP (binary).[4-6]
It was found that different barrier materials can lead to different degree of
lateral composition modulation, and the period of lateral
modulation ranges from 100\AA \, to 300 \AA \, depending on
the growth time and temperature.

	In this paper, we explore the usefulness 
of InGaAs quantum wires (QWR's) grown by the strain-induced lateral ordering 
(SILO) process for IR detection. 
Our theoretical modeling inculdes the effects of realistic band 
structures and microscopic strain distributions by combining the effective
bond-orbital model (EBOM) with the valence-force-field (VFF) model.
One of the major parameters for the IR detectors is the absorption 
quantum efficiency which is directly related to the absorption
coefficient by $\eta=1- e^{-\alpha l}$ where $\alpha$ is the absorption
coefficient and $l$ is the sample length. Thus, to have a realistic
accessment of the materials for device application, we need to
perform detailed calculations of the absorption coefficient, 
taking into account the excitonic and band structure effects. 
Both inter-subband and inter-band transitions
are examined systematically for a number of structure parameters 
(within the experimentally feasible range) chosen
to give the desired effect for IR detection. 

	It is found that the wavelengths for the
inter-subband transitions of InGaAs self-assembled QWR's range from 
10 to 20 $\mu m$,
while the inter-band transitions are around 1.5 $\mu m$. Thus, the material
provides simultaneneous IR detection at two contrasting wavelengths,
something desirable for application in multi-colored IR vedio camera.

Several structure models with varying degrees of alloy mixing for lateral 
modulation are considered. For the inter-band absorption, the excitonic
effect is important, since it gives rise a large shift in transition energy
and substantial enhancement of
the absorption spectrum.
To study the excitonic effect on the absorption spectrum for both discrete and
contunuum states,  we use a large set of basis functions with a finite-mesh
sampling in the k-space and diaginalize the exciton Hamiltonian directly.
Emphasis is put on the analysis of line shapes of various peak structures
arising from discrete excitonic states of one pair of subbands coupled
with the excitonic (discrete and continuum) states associated with other
pairs of subbands. We find that the excitonic effect enhances 
the first absorption peak around 1.5 times and shifts
the peak position by 20-30meV. 


\section {Theoretical model}

The QWR structures considered here consist of 8 pairs of
(GaAs)$_2$(InAs)$_{2.25}$ short-period superlattices (SPS) sandwiched between
Al$_{0.24}$Ga$_{0.24}$In$_{0.52}$As  barriers. 
The SPS structure prior to  strain induced lateral
ordering (SILO) is depicted in Fig. 1.
With lateral ordering, the structure is 
modeled by a periodic modulation of alloy composition
in layers with fractional monolayer of (In or Ga) in the SPS structure.
In layers 7 and 9 (starting from the bottom as layer 1), we have   
\begin{equation}
x_{\mbox{In}}=  \left\{ \begin{array}{ll}
 x_m [1-\sin (\pi y'/2b)] /2 &  \mbox{ for } y'< b\\ 
	0  & \mbox{ for } b < y' < L/2-b \\
 x_m \{1 + \sin [\pi (y'-L/2)/2b]\} /2 &  \mbox{ for } L/2-b < y' < L/2+b \\
 x_m   &	  \mbox{ for } L/2+b < y' < L-b \\
 x_m \{1 - \sin [\pi (y'-L)/2b]\} /2 &  \mbox{ for } y'> L-b, \end{array} \right.
\end{equation}
where $x_m$ is the maximum In composition in the layer, $2b$ denotes the 
width of lateral composition grading, and $L$ is the period of the
lateral modulation in the [110] direction. The experimental feasible range of
$L$ is between 100 \AA and 300 \AA. The length of $L$ is controled by
the growth time and temperature.

In layers 3 and 13, we have
\begin{equation}
x_{\mbox{In}}=  \left\{ \begin{array}{ll}
 0   &    \mbox{ for } 0 < y' < 5L/8-b \\
 x_m \{1 + \sin [\pi (y'-5L/8)/2b]\} /2 &  \mbox{ for } 5L/8-b<y'< 5L/8+b,\\ 
 	x_m  & \mbox{ for } 5L/8+b < y' < 7L/8-b \\
 x_m \{1 - \sin [\pi (y'-7L/8)/2b]\} /2 &  \mbox{ for } 7L/8-b < y' < 7L/8+b \\
 0   &    \mbox{ for } 7L/8+b < y' < L.
\end{array} \right.
\end{equation}
Similar equation for $x_{\mbox{Ga}}$ in layers 5 and 11 can be deduced
from the above.
By varying the parameters $x_m$ and $b$, we can get different degrees
of lateral alloy mixing. Typically $x_m$ is between 0.6 and 1, and $b$
is between zero and $15 a_{[110]} \approx$ 62 \AA.

A VFF model[13-15] is used to find the equilibrium 
atomic positions in the self-assembled QWR structure by minimizing the lattice energy.
The strain tensor at each atomic (In or Ga) site is then obtained by
calculating the local distortion of chemical bonds.

Once the microscopic strain distribution in the model structure is determined,
the energy levels and wave-functions of self-assembled 
quantum wires are then calculated within the effective bond-orbital model 
(EBOM). Detailed description of this method can be found in Refs. 
\cite{Cha}$^,$\cite{Mat}$^-$\cite{Osb}. 
EBOM used here is a tight-binding-like model in which two s-
like conduction bands (including spin) and four valence bands 
with total angular momentum $J=3/2$ (due to spin-orbit coupling
of p-like orbitals  with the spinor).  Thus, 
the present model is comparable to the six-band ${\bf k\cdot p}$
model as adpoted in Ref. \onlinecite{Ehe}

To minimize the computing effort, we express 
the electron and hole states for the quantum wire
structures in terms of eigen-states of a quantum well structure
with different in-plane wave vectors. 
The quantum well consists of 8 pairs of (GaAs)$_2$(InAs)$_2$ short-period
superlattice (SPS) plus two InAs monolayers (one inserted after the second
pair of SPS and the other after the sixth pair of SPS), so the the total
In/Ga composition ratio is consistent with the (GaAs)$_2$(InAs)$_{2.25}$ SPS.
The whole stack of SPS's is then sandwiched
between two slabs of Al$_{0.24}$Ga$_{0.24}$In$_{0.52}$ barriers.  
Let us denote the quantum well eigen-states 
as $|n,k_1,k_2\rangle_{QW}$ where $n$ labels the subband, 
$k_1$ denotes the wave vector
along the wire ($[1 \bar 1 0]$) direction 
and $k_2$ labels the wave vector in the [110]
direction, which is perpendicular to the
wire and the growth axis. Expanding the quantum well states in terms of
bond-orbitals, we have
\[ |n,k_1,k_2\rangle_{QW}=\frac 1 {\sqrt{L}} \sum_{\alpha,{\bf R}}
f_{n,k_1,k_2}(\alpha,R_z)\exp(ik_2 R_2)
\exp(ik_1 R_1)|u_{\alpha}({\bf R})\rangle, \]
where $L$ is the sample length along the wire axis, 
$f_{n,k_1,k_2}(\alpha,R_z)$ is the eigen-vector for the quantum well
Hamiltonian and $u_{\alpha}({\bf R})$ denotes
an $\alpha$-like bond orbital state at site ${\bf R}$ ($\alpha=1,\cdots,6$
for two s-like conduction-band and four $J=3/2$ 
valence-band orbitals). Here ${\bf R}$ runs over all lattice sites within
the SPS layer (well region) and AlGaInAs  layer (barrier region).

We  then diagonalize
the hamiltonian for the quantum wire (QWR) within a basis which consists of
the quantum well states with $k_2$'s separated by reciprocal lattice vectors
$g_m = m (2\pi/a_{[110]})$; $m$= integers. 
Namely, 
\[ |i,k_1,k_2\rangle= \sum_{n,m} C_{i,k_1}(n,k_2+g_m)
|n,k_1,k_2+g_m\rangle_{QW} \]
where $C_{i,k_1}(n,k_2+g_m)$ is the eigen-vector for the 
quantum-wire hamiltonian
matrix for the $i-th$ QWR subband at wave vector $(k_1,k_2)$.
 
In terms of the bond orbitals, we can rewrite the QWR states as

\[ |i,k_1,k_2 \rangle=\sum_{\alpha,{\bf R}} F_{i,k_1,k_2}
(\alpha,{\bf R}) |u_{\alpha}({\bf R})\rangle \]
where
\[ F_{i,k_1,k_2}(\alpha,{\bf R})=\frac 1 {\sqrt{L}} 
\sum_{n,m} C_{i,k_1}(n,k_2+g_m)f_{n,k_1,k_2+g_m}(\alpha,R_z)
\exp[i(k_2+g_m)R_2] \exp(ik_1 R_1) \]
is the QWR envelope function.
For the laterally confined states, the dispersion of bands versus $k_2$
is negligible; thus, the $k_2$ dependence can be ignored.

The absorption coefficeient for inter-subband transitions between 
subbands $i$ and $j$ is given by
\begin{equation}
 \alpha_{ij}(\hbar\omega)=\frac{4\pi^2e^2\hbar}{n_r~ m_e^2~c~V~\hbar\omega}
\sum_{k_1,k_2}|\langle i,k_1,k_2|\hat{\epsilon}
\cdot {\bf p}|j,k_1,k_2\rangle|^2
[f_j(k_1,k_2)-f_i(k_1,k_2)] \delta(E_j(k_1,k_2)-E_i(k_1,k_2)-\hbar\omega) 
\end{equation}
where $n_r$ is the refractive index of the 
QWR, $V$ is the volume of the QWR sample restricted within the SPS region,
$f_i(f_j)$ is the Fermi-Dirac distribution function for
subbnad $i$ ($j$). The optical matrix elements between QWR subband states
are related to those between bond orbitals by 
\[ \langle i,k_1,k_2|\hat{\epsilon} \cdot {\bf p}|j,k_1,k_2\rangle = 
\sum_{\alpha,\alpha',\tau} F^*_{i,k_1,k_2}(\alpha,{\bf R}) 
F_{j,k_1,k_2}(\alpha',{\bf R})
\langle u_{\alpha}({\bf R})|\hat{\epsilon}\cdot {\bf p}|
u_{\alpha'}({\bf R}+\vec \tau) \rangle, \]
where $\vec \tau$ runs over on-site or the 12 nearest-neighbor sites 
in the fcc lattice.
The optical matrix elements between bond orbitals are related to the 
band parameters by requiring the optical matrix elements between bulk
states near the zone ceneter to be identical to those obtained in
the ${\bf k \cdot p}$ theory\cite{kdp}. We obtain\cite{cit}
\[ langle u_{s}({\bf R})|p_{\alpha}
u_{\alpha'}({\bf R}) \rangle = \sqrt{m_0 E_p/2}
\delta_{\alpha,\alpha'}; \alpha=x,y,z, \]
\[ langle u_{s}({\bf R})|p_{\alpha}
u_{s}({\bf R}+\vec \tau) \rangle = (\hbar/2\sqrt{2}a)(E_p/Eg-m_0/m_e^*) 
\tau_{\alpha} ; \alpha=x,y,z, \]
where $\tau_{\alpha}$ is the $\alpha$-th of the lattice vector $\tau$
in units of a/2, $E_p$ is the inter-band optical matrix element as defined in
Ref. \onlinecite{kdp}, and $m_e^*$ is the electron effective mass.

Next, we study the inter-band transitions. For this case, the excitonic
effect is important. Here we are only interested in the absorption spectrum
near the band edge due to laterally confined states. Thus, the dispersion
in the $k_2$ direction can be ignored. The exciton
states with zero center-of-mass momentum
can then be written as linear combinations of products of
electron and holes states associated with the same $k_1$
(wave vector along the wire direction). We write the electron-hole
product state for the $i$-th conduction
subband and $j$-th valence subband as

\[ |i,j;k_1\rangle_{ex}=|i,k_1\rangle |j,k_1\rangle\]
\[\equiv \sum_{\alpha,\beta,{\bf R}_e,{\bf R}_h}F_{i,k_1}(\alpha,{\bf R}_e)G_{j,k_1}
(\beta,{\bf R}_h)|u(\alpha,{\bf R}_e)>|u(\beta,{\bf R}_h)>.\]
               
The matrix elements of the exciton Hamiltonian within this basis is given by
\begin{equation}
   \langle i,j,k_1|H_{ex}|i',j',k_1'\rangle = 
[E_i(k_1)\delta_{i,i'}-E_j(k_1)\delta_{j,j'}]
- \sum_{{\bf R}_e,{\bf R}_h} {\cal F}_{ii'}^*({\bf R}_e)v({\bf R}_e
-{\bf R}_h){\cal G}_{jj'}({\bf R}_2),\end{equation}
where $v({\bf R}_e,{\bf R}_h)=\frac{4\pi~e^2}{\epsilon(0)~|{\bf R}_e-
{\bf R}_h|}$ is
the coulomb interaction between the electron and hole screened by
the static dielectric constant $\epsilon(0)$, and
\[ {\cal F}_{ii'}({\bf R}_e) = \sum_{\alpha}F^*_{i,k_1}(\alpha,{\bf R}_e)
F_{i',k_1}(\alpha,{\bf R}_e) \]
describes the charge density matrix for the electrons.
Similarly, 
\[ {\cal G}_{jj'}({\bf R}_h) = \sum_{\beta}G^*_{j,k_1}(\beta,{\bf R}_h)
G_{j',k_1}(\beta,{\bf R}_h) \]
describes the charge density matrix for the holes.
In Eq. (x), we have adopted the approximation
\[   \langle u(\alpha,{\bf R}_e)|\langle u(\beta,{\bf R}_h)|v
|u(\alpha',{\bf R}'_e)\rangle|u(\beta',{\bf R}'_h)\rangle             
\approx v({\bf R}_e-{\bf R}_h)\delta_{\alpha,\alpha'}\delta_{\beta,\beta'}
\delta_{{\bf R}_e,{\bf R}'_e} \delta_{{\bf R}_h,{\bf R}'_h}, \]
since the Coulomb potential is a smooth function over the distance of 
a lattice copnstant, except at the origin, and the bond orbitals are
orthonormal to each other.
At the origin (${\bf R}_e={\bf R}_h$), the potential is singular, and
we replace it by an empirical constant which is adjusted so as to give
the same exciton binding energy
as obtained in the effective-mass theory for a bulk
system. The results are actually insensitive to the on-site Coulomb
potential parameter, since the Bohr radius of the exciton is much larger
than the lattice constant.

After the diagonalization, we obtain the excitonic states as linear 
combinations of the electron-hole product states, and the 
inter-band absorption coefficient is computed according to

\begin{equation}
\alpha^{ex}(\hbar\omega)=\frac{4\pi^2e^2\hbar}{n_r~ m_e^2~c~A~\hbar\omega}
\sum_{n}|\sum_{k_1,i,j}C^n_{ij}(k_1)\hat{\epsilon}\cdot {\bf p}_{ij}(k_1)|^2\delta(
\hbar\omega-E^{ex}_{n}),
\end{equation}
where $A$ is the cross-sectional area of the SPS region within
the QWR unit cell (as depicted in Fig. 1). $C^n_{ij}(k_1)$ is the
$n$-th eigen-vector obtained by diagonalizing the exciton Hamiltonian of Eq.
(4). The momentum matrix elements ${\bf p}_{ij}(k_1)$ are given by 

\begin{equation}
{\bf p}_{ij}(k_1)=
\sum_{\alpha,\beta,\vec \tau}F^*_{i,k_1}(\alpha,{\bf R}_e) 
G_{j,k_1}(\beta,{\bf R}_h+\vec \tau)
\langle u_{\alpha}({\bf R}_e)|\hat{\epsilon}\cdot {\bf p}|
u_{\beta}({\bf R}_h+\vec \tau) \rangle. \end{equation}
In the absence of band mixing, the conduction subband state reduces to a pure
$s$-like state and the valence subband state reduces to a pure $p$-like state.
In that case, only the element $\langle u_{s}({\bf
R}_e)|p_{\alpha}|u_{p_{\alpha}}({\bf R}_e) \rangle = \sqrt{m_0E_p/2} $ 
is needed,

In order to obtain a smooth absorption spectrum, we replace the $\delta$ 
function in Eq. (1)
by a Lorentzian function with a half-width $\Gamma$, 

\begin{equation}
\delta(E_i-E)\approx \Gamma/\{\pi[(E_i-E)^2+\Gamma^2]\}
\end{equation}

$\Gamma$ is energy width due to imhomogeneous broadening, which is taken to be 
0.01 eV (??).

\newpage

\section {Results And Discussions}

       We have performed calculations of inter-subband and inter-band 
absorption spectra for
the QWR structure depicted in Fig. 1 with varaying degree of alloy mixing and
different lengths of period ($L$) in lateral modulation. We find that
the inter-subband absorption
spectra are sensitive to the length of period ($L$), but rather insensitive to
the degree of of alloying mixing. Thus, we only present results for
the case with moderate alloy mixing, which are characterized by parameters
$b=33\AA$ and $x_m=1.0$. In all the calculations, the bottom layer atoms of QWR's
are bounded by the InP substrate, while the upper layer atoms and GaAS capping 
layer atoms are allowed to move freely. This strucure is corresponding to the
unclamped struture as indicated in reference \cite{li1}.
	
	For different period length L of QWR's, the strain distribtion profiles 
are qualitatively similar as shown in reference \cite{li1}. As L decreases, the
hydrostatic strain in rich In region (i.e. right half zone of QWR's unit) increase, 
while it decreases in rich Ga region. The bi-axial strain has the opposite change with
L. The variation of hydrostatic and bi-axial strains with deducing QWR's period reflects in
the potential profiles as the difference of CB and VB band eage increases, which can be
seen in Figure 2. 

	It can be easily understood that the shear strains increase when L is decuded.

	The potential profiles due to strain-induced lateral ordering
seen by an electron in two  QWR structures considered here ($L=50 a_{[110]}$
and $L=40 a_{[110]}$) are shown in Fig. 2. more discussions...

	The conduction subband structures for the self-assembled QWRs with alloying mixing
($x_m=1.0$ and $b=8a_{[110]}$) for ($L=50 a_{[110]}$
and $L=40 a_{[110]}$) are shown in Fig. 3. All subband are grouped
in pairs with a weak spin splitting (not resolved on the scale shown).
For $L=50 a_{[110]}$, the lowest three pairs of subbands are nearly
dispersionless along the $k_2$ direction, indicating the effect of
strong lateral confinement. The inter-subband transition between the first
two pairs give rise to the dominant IR response at photon energy around
60 meV.  For $L=40 a_{[110]}$, only the lowest pair of subbands (CB1) is 
laterally confined (with a weak $k_2$ dispersion). The higher subbands 
corresponding to laterally unconfined states (but remain confined along
the growth axis) and they have large dispersion versus $k_2$. 
We find three pairs
of subbands (CB2-CB4) are closely spaced in energy (within 5 meV?). State
orgion of degeneracry?? 
 
	The valence subband structures for the self-assembled QWRs with alloying mixing
($x_m=1.0$ and $b=8a_{[110]}$) for ($L=50 a_{[110]}$
and $L=40 a_{[110]}$) are shown in Fig. 4. more discussions??

\vspace{2ex}
{\bf  A. Inter-subband absorption }
\vspace{2ex}

	Inter-subband absorption
spectrum is the most relavent quantity in determining
the usefulness of self-assembled QWR's for application in IR
detection. Fig. 5 shows the calculated inter-subband absorption spectra
of the self-assembled QWR structure (as depicted in Fig. 1) for three different lengths
of period: $L=72, 50$, and $40 a_{[110]}$ (approximatley 300 \AA, 200 \AA, and
160 \AA, respectively). 
In the cacluation, we assume that these QWR structures are n-type
doped with linear carrier density around $1.65 \times 
10^6cm^{-1}$ (which corresponds to
a Fermi level around 25 meV above the condunction band minimum).

For comparison purposes, we show results for
polarization vector along both the [110] (solid curves)
and [001] directions (dashed curves). The results for [1$\bar 1$0]
polarization are zero due to the strict translational invariance imposed in our
model calculation.  

The  peak positions for the
inter-subband transition with normal incidence (with [110] polarization) are 
around 65 meV, 75 meV, and  110 meV for the three cases considered
here. All these are within the desirable range of IR detection. 
As expected, the transition energy increases as the length of 
period decreases due to the increased degree of lateral confinement. 
However, the transition energy will saturate at around 110 meV as we
further reduce the length of period, since the bound-to-continuum
transition is already reached at $L=40 a_{[110]}$.

The absorption strengths for the first two cases ($L=72 a_{[110]}$
and $L=50 a_{[110]}$) are reasonably strong (around 400 $cm^{-1}$ and
200 $cm^{-1}$, respectively). They both correspond to the bound-to-bound
transitions. In contrast, the absorption strength for the third case
is somewhat weaker (around 50 $cm^{-1}$), since it corresponds to the
bound-to-continuum transition. For comparison, the absorption strength
for typical III-V QWIPs is around ?? 

The inter-subband absorption for the [001] polarization is peaked around
?? meV. The excited state involved in this transition is a quantum
confined state due to the Al$_{0.24}$Ga$_{0.24}$In$_{0.52}$ barriers.
Thus, it has the same physical origin as the inter-subband
transition used in typical QWIP structure.
Although this peak is not useful for IR detection with normal
incidence, it can be used as the second-color detection
if one puts a diffraction grating on the surface as typically done 
in the fabrication of QWIPs.

\vspace{4ex}
{\bf B. Inter-band absorption }
\vspace{2ex}

	The inter-band optical transitions are important for the
characterization of self-assembled QWR's, since they are readily observable via
the Photoluminescence (PL) or optical transimission experiment.
For IR-detector application, they offer another absorption peak at
mid IR wavelengths, which can be used together with the inter-subband
transitions occured at far IR wavelengths for multi-colored detection.
Thus, to understand the full capability of the self-assembled QWR material, we
also need to analyze the inter-band absorption.

   Fig. 6 shows the squared optical matrix elements versus $k_2$ 
for two self-assembled QWR's considered in the previous section (with $L=50$ and $40
a_{[110]}$). For the case with $L=50 a_{[110]}$, the optical matrix elements
for both [110]
and [1$\bar 1$0] polarizations are strong with a polarization
ratio $P_{[1\bar 1 0]}/P_{[110]}$ around 2. This is similar to the case
with $L=72 a_{[110]}$ as reported in Ref. xx.
For the case with $L=40 a_{[110]}$, the optical matrix elements
for both [110]
and [1$\bar 1$0] polarizations are weak. This is due to the fact that
the electrons and hole are laterally confined in different regions
in the QWR, as already indicated in the potential profile as shown in
Fig. 2(b). Thus, the inter-band absorption for this case will be uninteresting.

	Fig. 7 shows the inter-band absorption spectra for SILO
QWR's with $L=72$ and $50 a_{[110]}$, including the excitonic effects. 
The PL properties of the $L=72 a_{[110]}$ structure with alloying mixing
characterized by $x_m=0.1$ and $b=8a_{[110]}$ has been studied
in our previous paper. The QWR structure has a gap around 0.74 eV
with a PL polarization ratio ($P_{[1\bar 10]}/P_{[110]}$) around 3.1. 
The absorption coefficient for this structure has
a peak strength around 250 cm$^{-1}$. The binding energy
for the ground state exciton labeled 1-1 (derived primarily from
the top valence subband and the lowest
conduction subband) is around 20 meV. Thus, the peak position in the
absorption spectrum shifts from
0.76 meV (without the excitonic effect) to 0.74 meV 
(with the excitonic effect). The exctionic effect also 
enhances the peak strength from 200 cm$^{-1}$ to 250 cm$^{-1}$.
The other peak structures (labeled 2-2, 2-3,... etc.)
are derived primarily from the transitions between
the lower valence subbands to the higher conduction subbands). 

For the QWR structure with $L=50 a_{[110]}$, we obtain
similar absorption spectrum with a peak strength
around 400 $cm^{-1}$(??). The exciton binding is around 40?? meV, and the 
excitonic enhancement factor of the first peak is around 1.15 (??),
higer than the case with $L=72 a_{[110]}$. 
This indicates that the case with $L=50 a_{[110]}$ has
stronger lateral confinement for electrons and holes, which leads 
larger exciton binding energy and stronger excition oscillator strength
(due to the larger probability that the electron and hole
appear at the same position).
The secondary peaks due to excitonic states derived from higher
subbands are also subtantially stronger than their counterparts
in the $L=72 a_{[110]}$ case.

\newpage

\section {Summary and Discussions}

\vspace{2ex}

We have studied the inter-subband and inter-band
absorption spectra for self-assembled InGaAs quantum wires
for consideration in IR-detector application. 
Detailed band structures, microscopic strain distributions, and excitonic
effects all have been taken into account.
A number of realistic structures grown via strain-induced lateral 
ordering process are
examined. We find that the self-assembled InGaAs quantum wires 
are good candidate for multi-colored IR detector materials.
They offer two groups of strong IR absorption peaks: one in the 
far-IR range with wavelengths covering 10 - 20 $\mu m$ (via
the inter-subband transition), the other in the mid-IR range with
wavelengths centered around 1.5 $\mu m$ (via the inter-band
transition). Due the strain induced lateral modulation, 
the inter-subband transition
is strong for normal incident light with polarization along
the direction of lateral modulation ([110]). This  gives the
self-assembled InGaAs quantum wires a distinct advantage over the
quantum well systems for application in IR detection.

	The inter-subband absorption is found to be sensitive
to the length of period ($L$) of laterial modulation with the aborption
peak position varying from 60 meV to 110 meV as the length of period
is reduced from 300 \AA \, to 160 \AA. However, further reduction
in the length of period does not shift the absorption peak very much, as
the excited states become laterally unconfined.

	For the inter-band transition, we find that the excitonsic effect
enhances the absorption peak strength by about 10-20 \%,
and shift the peak position
by about  20-40 meV for the structures considered. 
The reduction in the period length ($L$) leads to 
stronger lateral confinement, hence larger exciton
binding and stronger absorprtion strength.
As conclusion, this paper should give the experiment the realistic guidance in
the growth of the IR detector and present the interesting physical thoughts
for the theoretists and experimentists.

In conclusion, we successfully demonstrated that self-assembled quantum wires
are promising IR-detector materials and we provided theoretical modeling for
the optical characteristics for realistic QWR structures, which can be used to
guide future fabrication of quantum wire infrared detectors.

\newpage
\vspace{2cm}
Figure Captions

\vspace{1ex}

\singlespace

\noindent Fig. 1.  Schematic sketch of the unit cell of
the self-assembled quantum wire for the model structure considered. 
Each unit cell consists of 8 pairs of (2/2.25) GaAs/InAs 
short-period superlattices (SPS).
In this structure, four pairs of (2/2.25) SPS (or 17 diatomic layers) form
a period, and the period is repeated twice in the unit cell.  
Filled and open circles indicate Ga and In rows 
(each row extends infinitely along the $[1\bar 1 0]$ direction). 

\noindent
Fig. 2. Conduction band and valence band edges for
self-assembled QWR structure depicted in Fig. 1
for (a) $L=50 a_{[110]}$ and (b) $L=40 a_{[110]}$. Dashed: without alloy
mixing. Solid: with alloy mixing described by $x_m=1.0$ and $b=8a_{[110]}$.

\noindent
Fig. 3.  Conduction subband structure of 
self-assembled QWR for (a) $L=50 a_{[110]}$ and  (b) $L=40 a_{[110]}$
with $x_m=1.0$ and $b=8a_{[110]}$.

\noindent
Fig. 4.  Valence subband structure of 
self-assembled QWR for (a) $L=50 a_{[110]}$ and  (b) $L=40 a_{[110]}$
with $x_m=1.0$ and $b=8a_{[110]}$.

\noindent
Fig. 5.  Inter-subband absorption spectra of 
self-assembled QWR for (a) $L=72 a_{[110]}$, (b) $L=50 a_{[110]}$, and  
(c) $L=40 a_{[110]}$ with $x_m=1.0$ and $b=8a_{[110]}$. 
Solid: [110] polarization, dashed : [001] polarization.

\noindent
Fig. 6.  Inter-band optical matrix
elements squared versus $k_1$ of self-assembled QWR's for 
(a) $L=50 a_{[110]}$ and  (b) $L=40
a_{[110]}$ with $x_m=1.0$ and $b=8a_{[110]}$.

\noindent
Fig. 7.  Inter-band absorption spectra of
self-assembled QWR's for (a) $L=72 a_{[110]}$ and (b) $L=50 a_{[110]}$ 
with $x_m=1.0$ and $b=8a_{[110]}$. 
Solid: [110] polarization with excitonic effect. 
Dotted : [1$\bar 1$0] polarization with excitonic effect. 
Dashed: [110] polarization without excitonic effect.

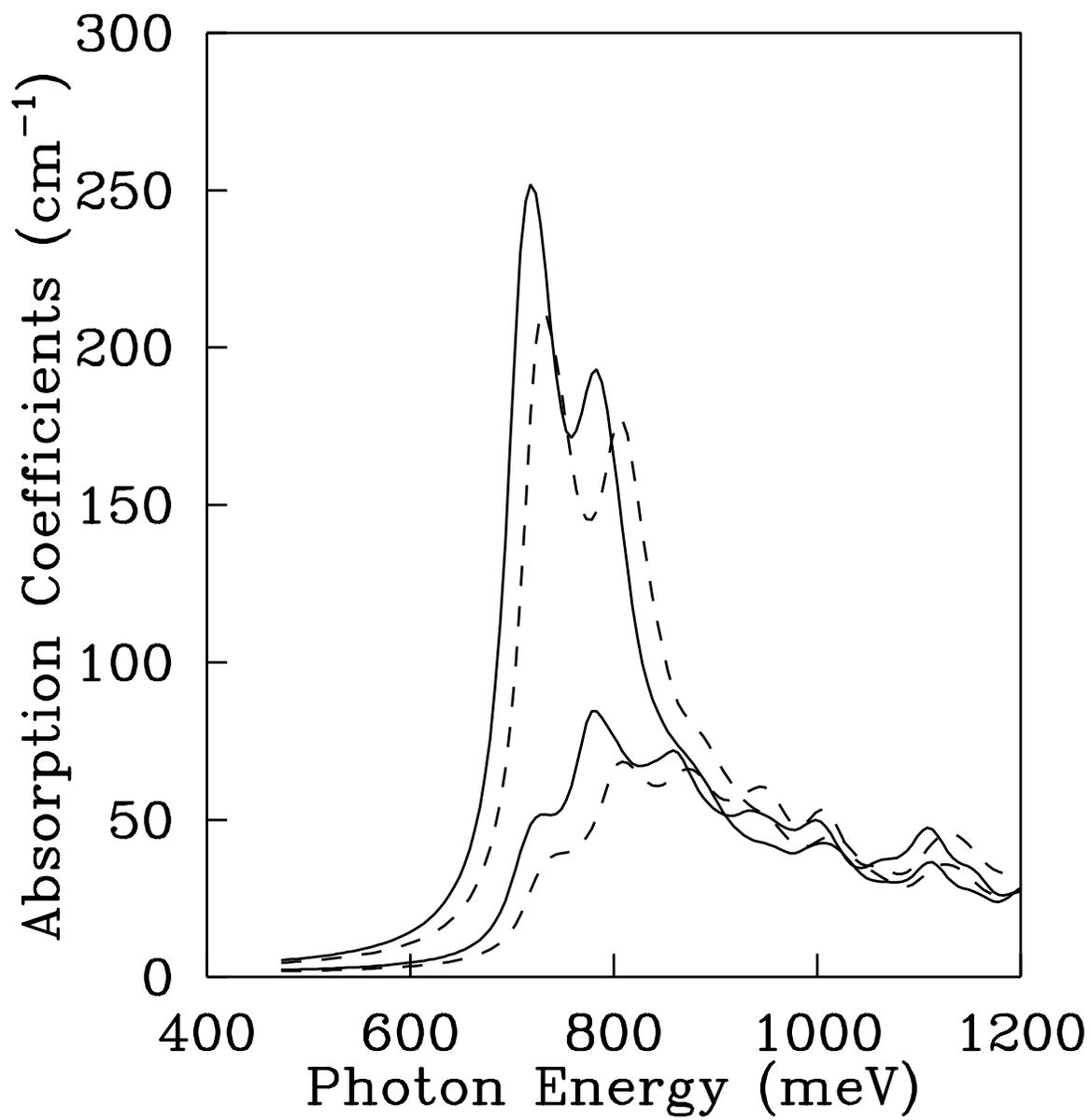
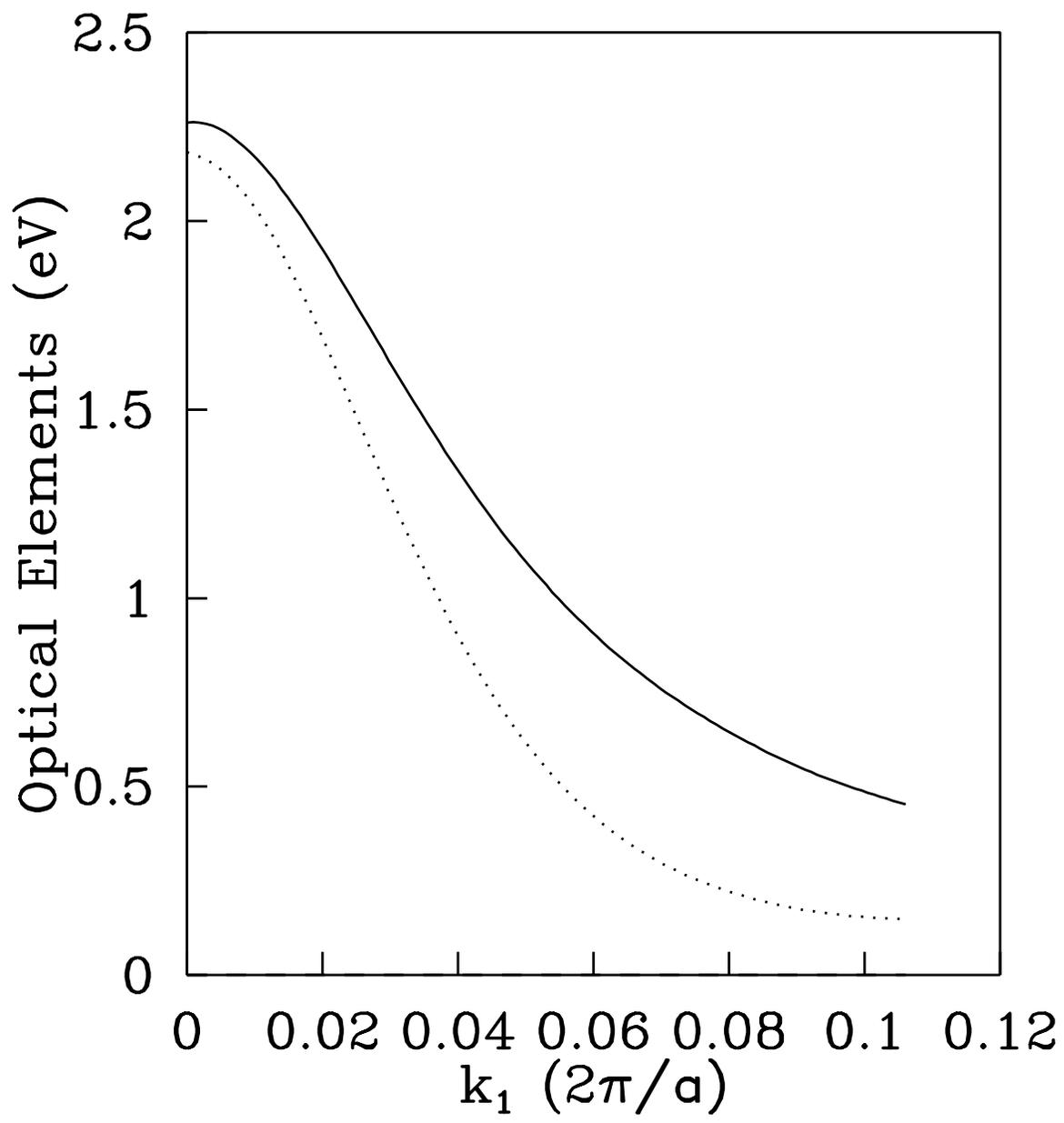
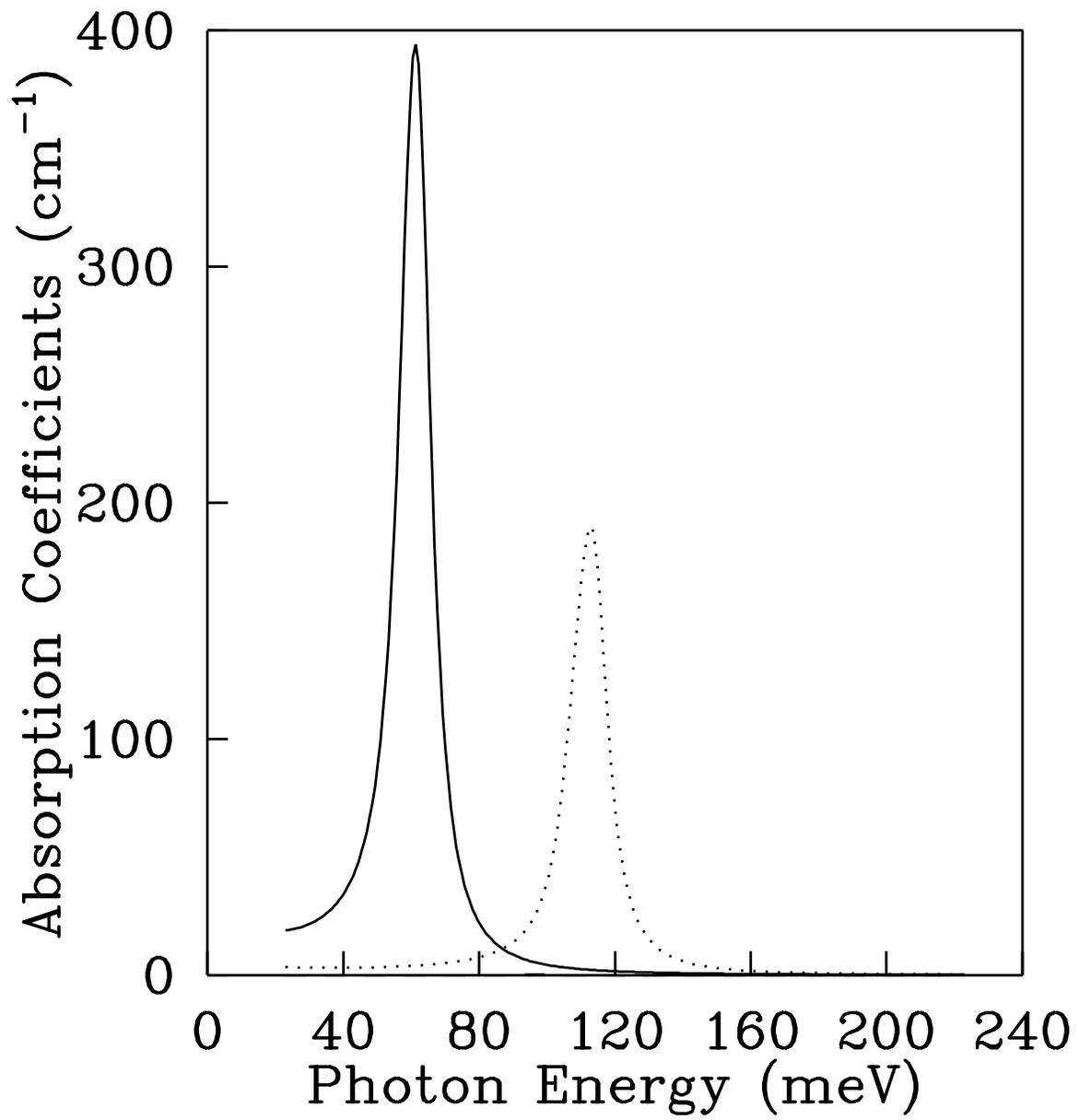
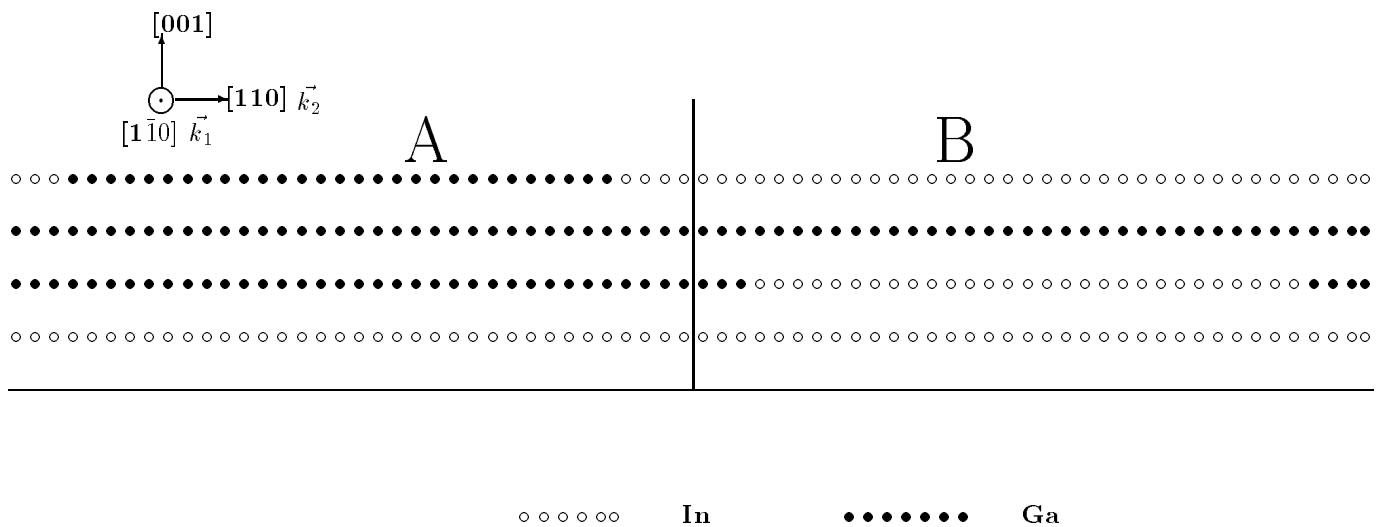
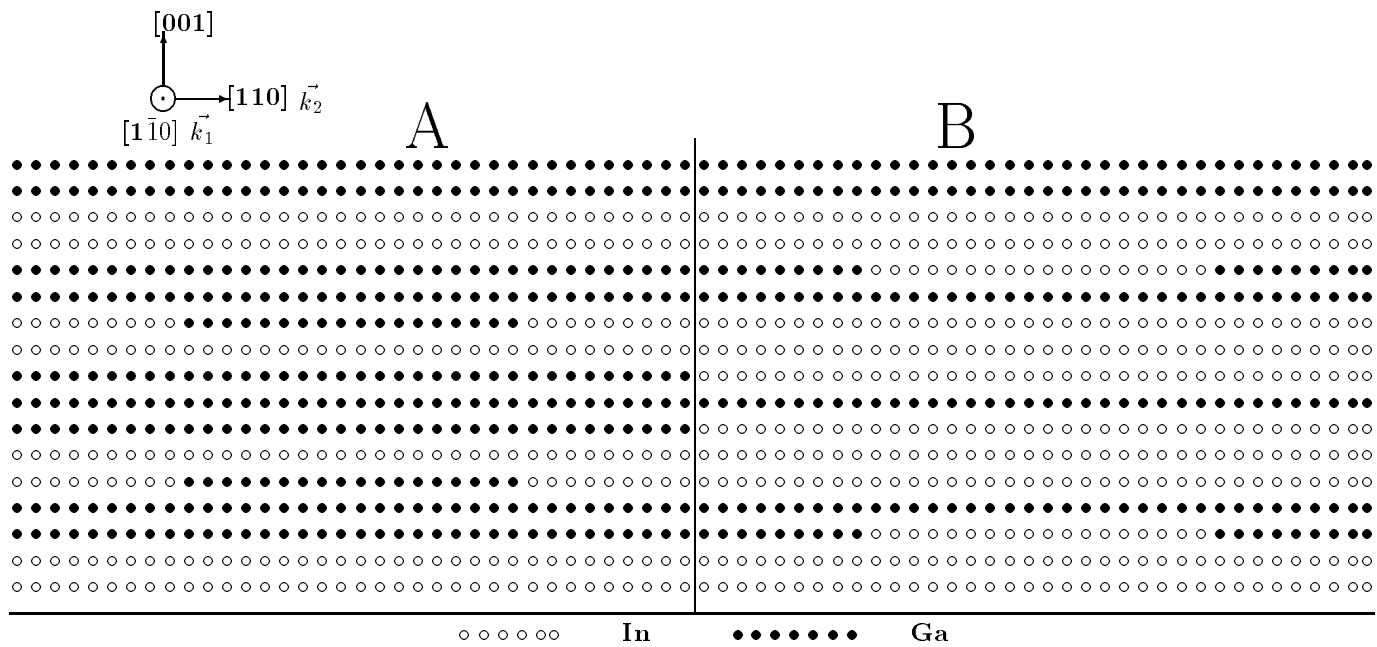
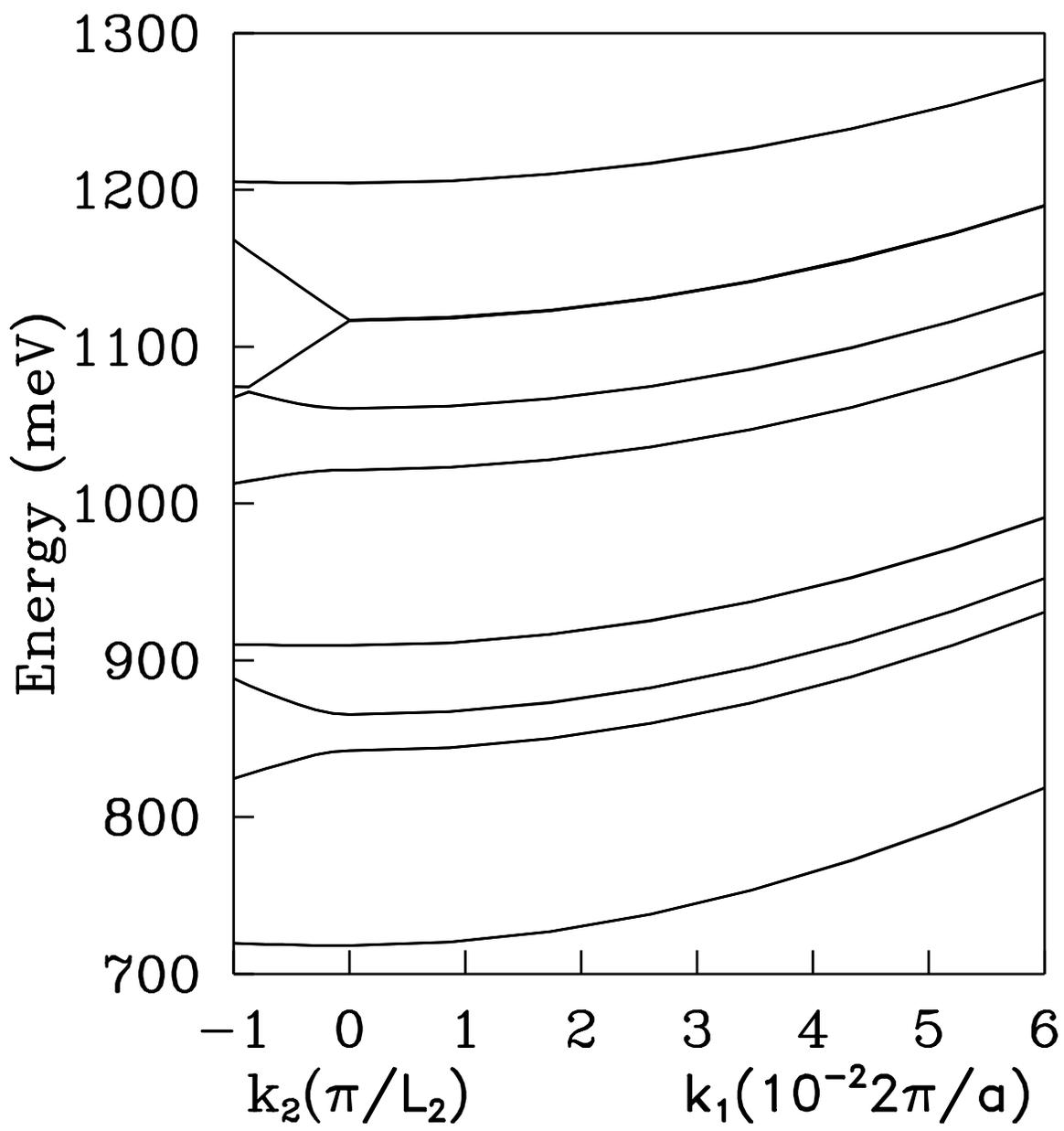
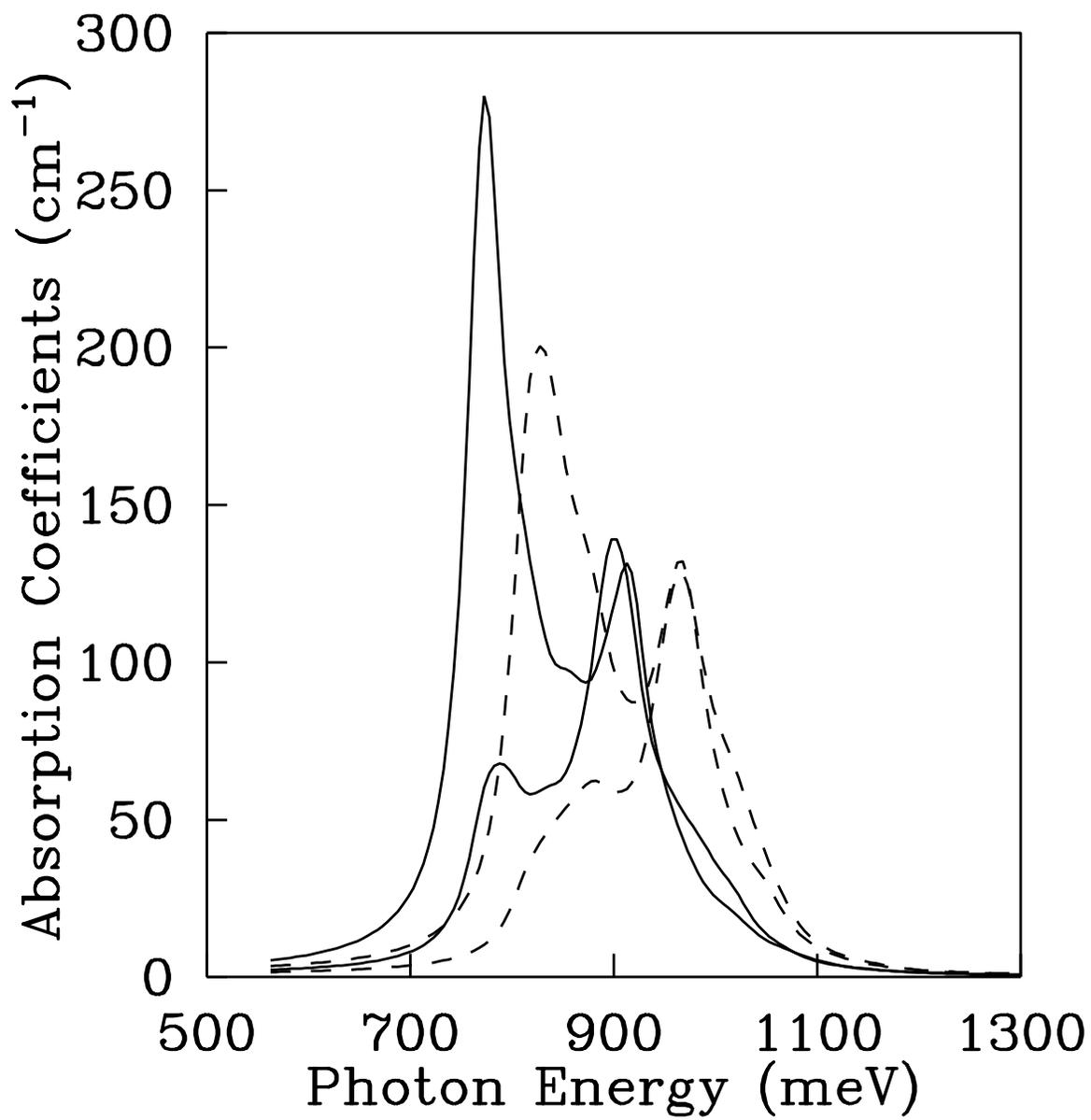
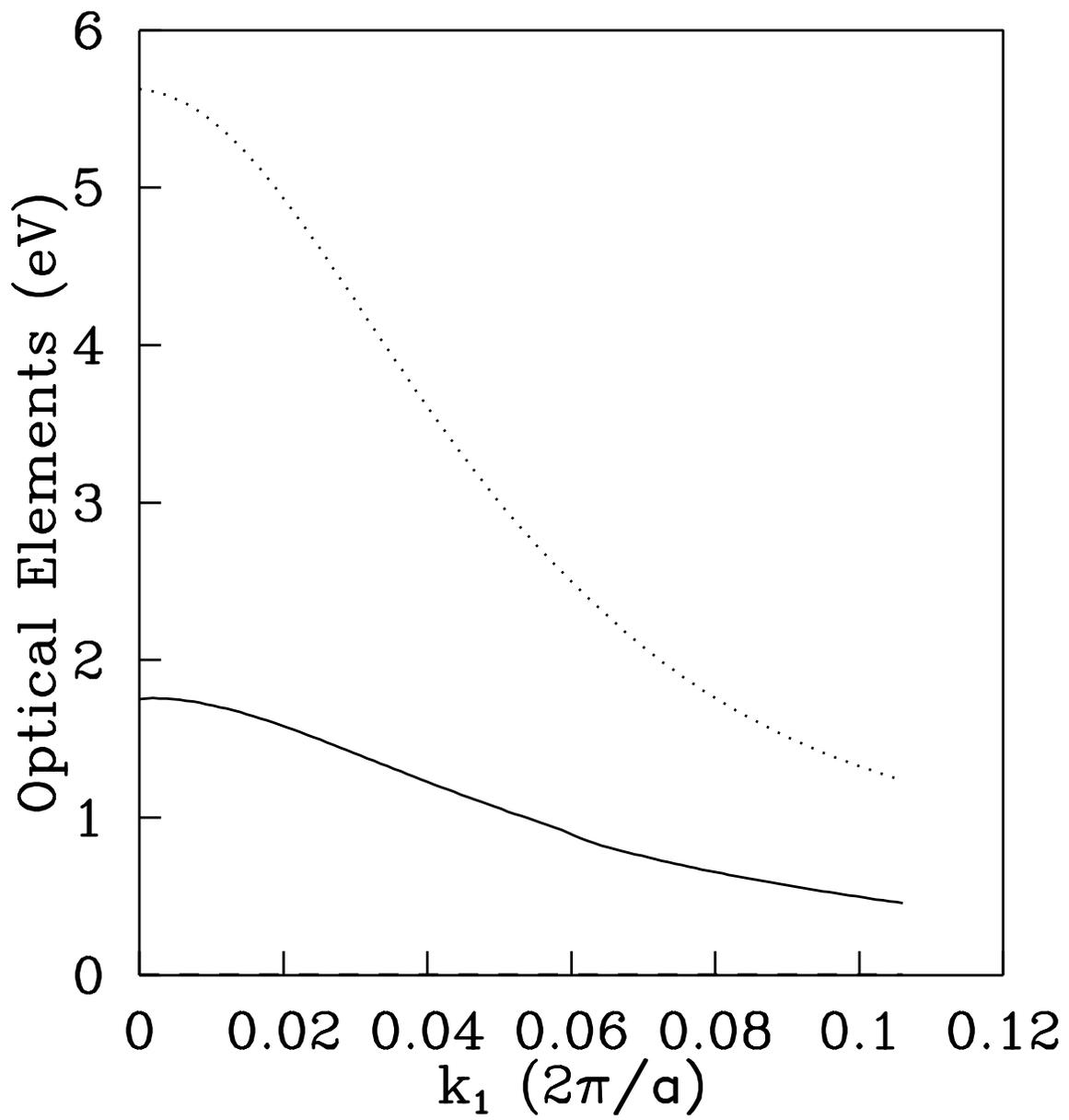
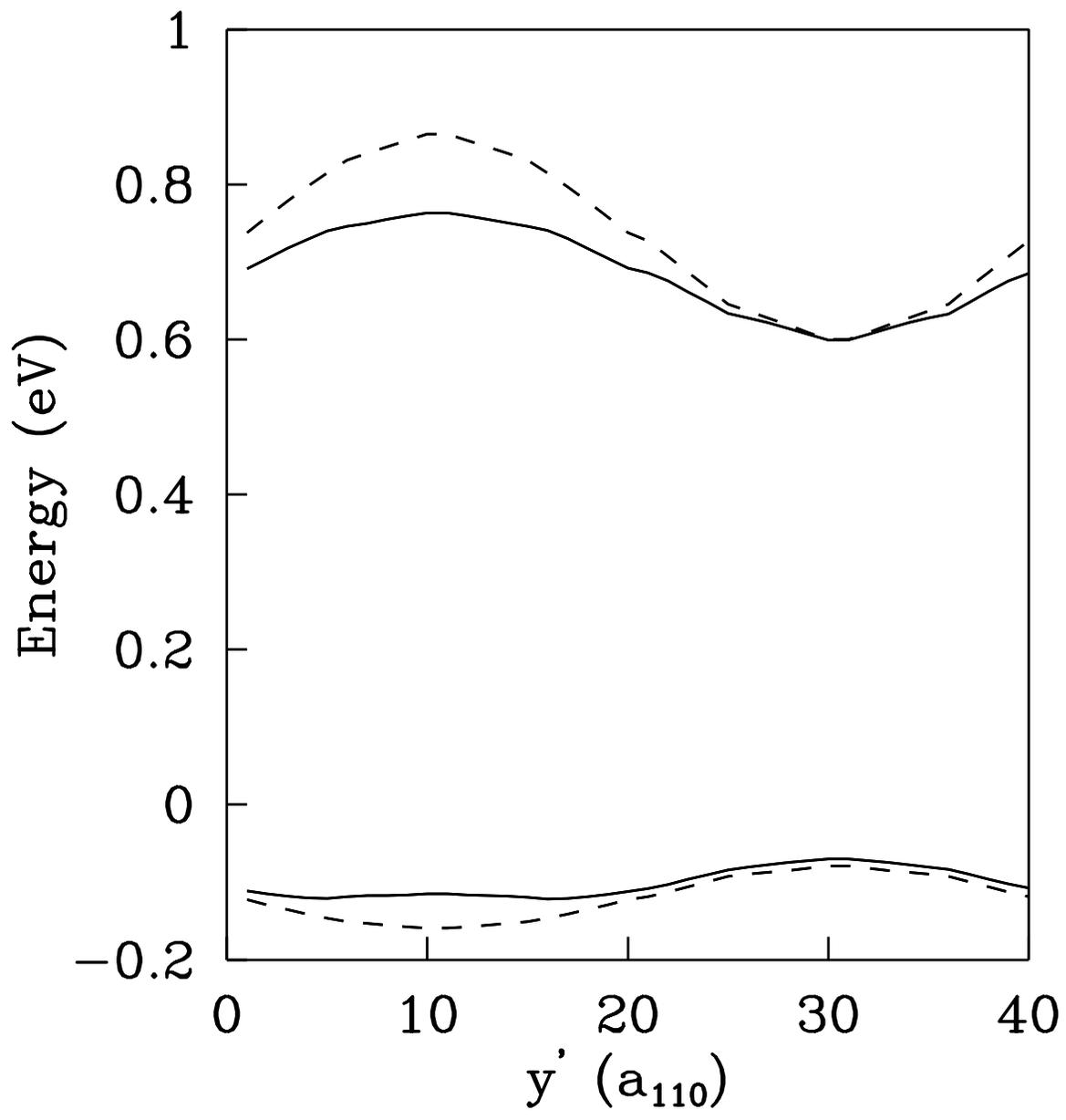
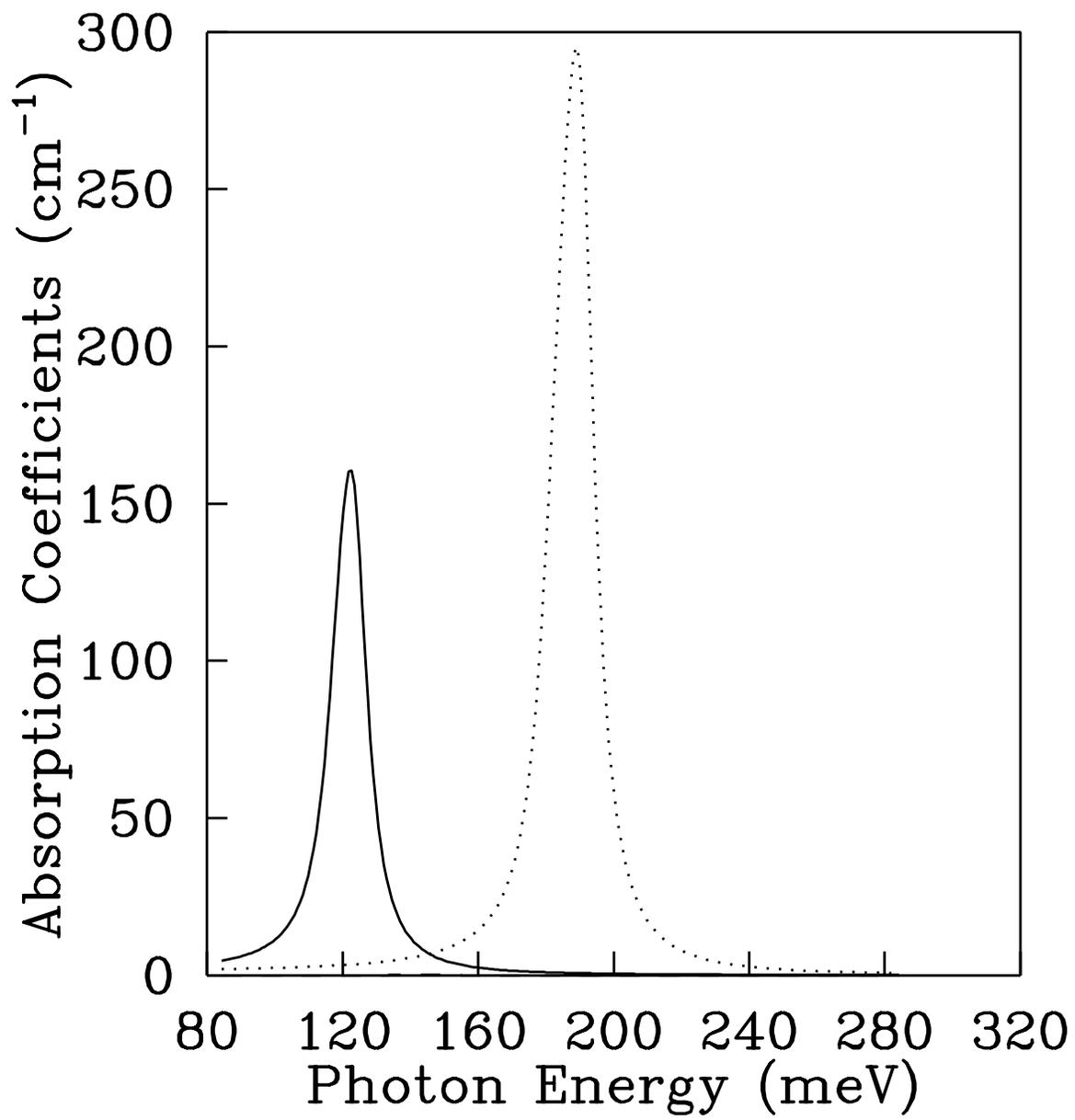
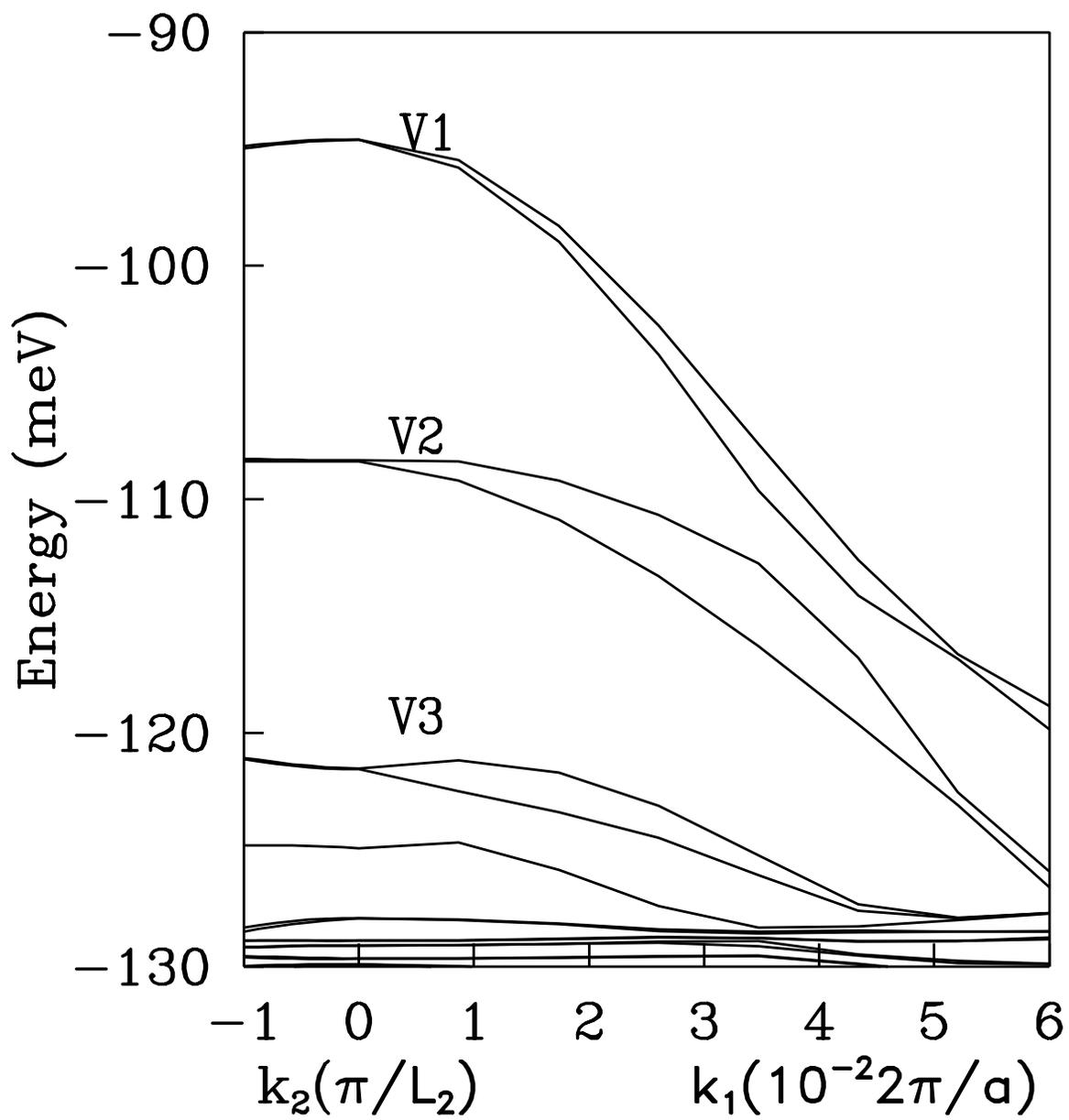
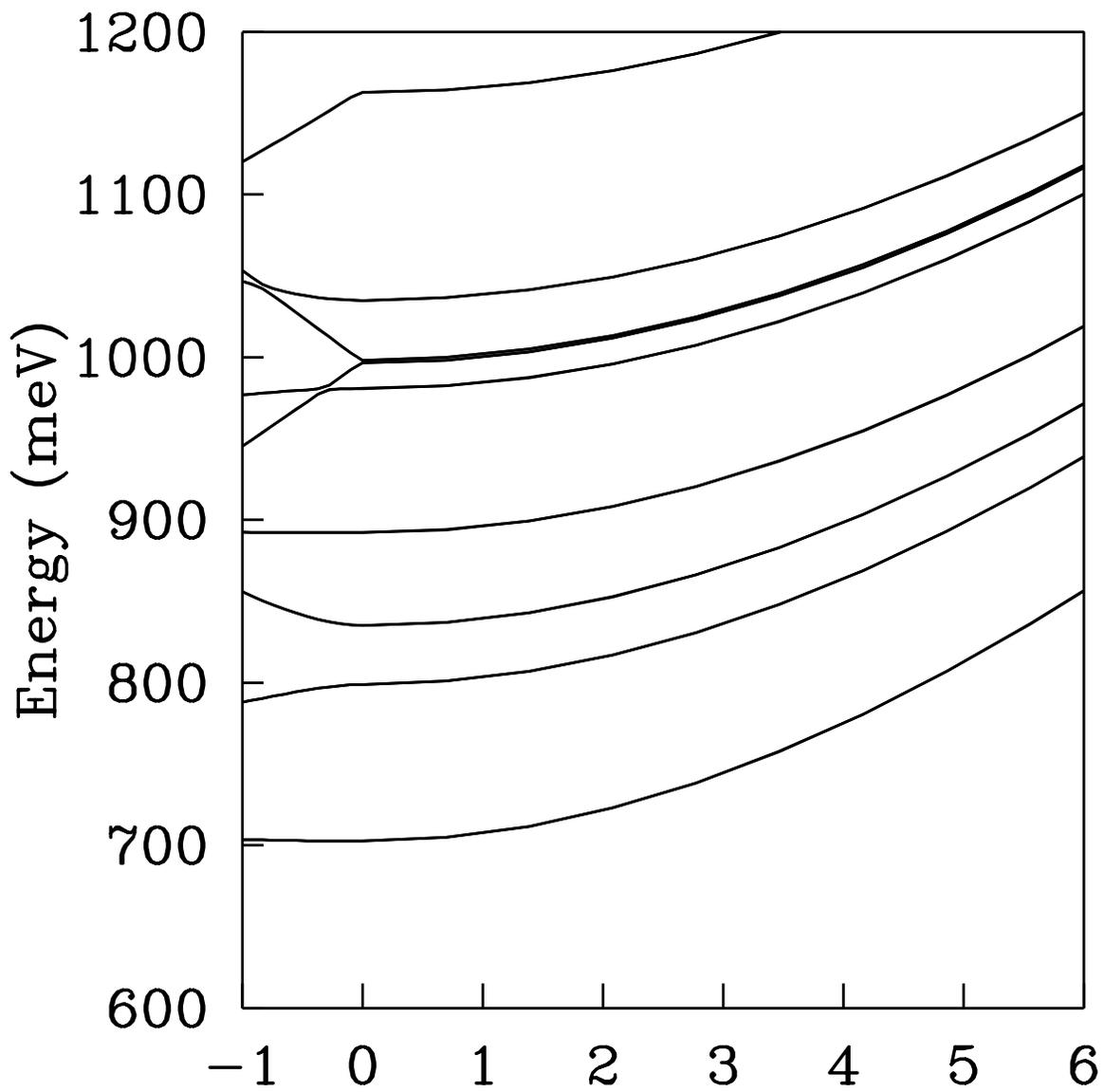
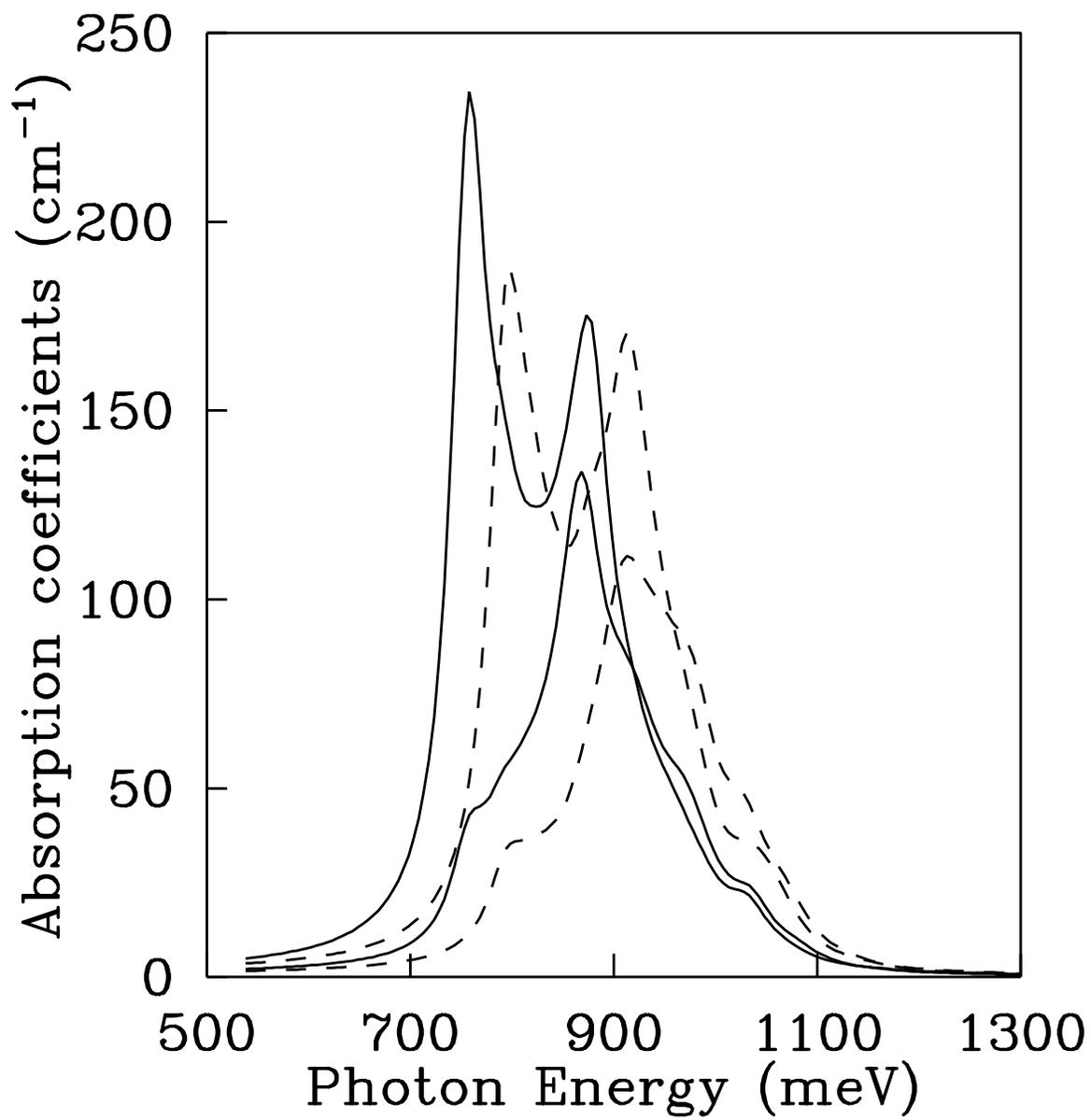
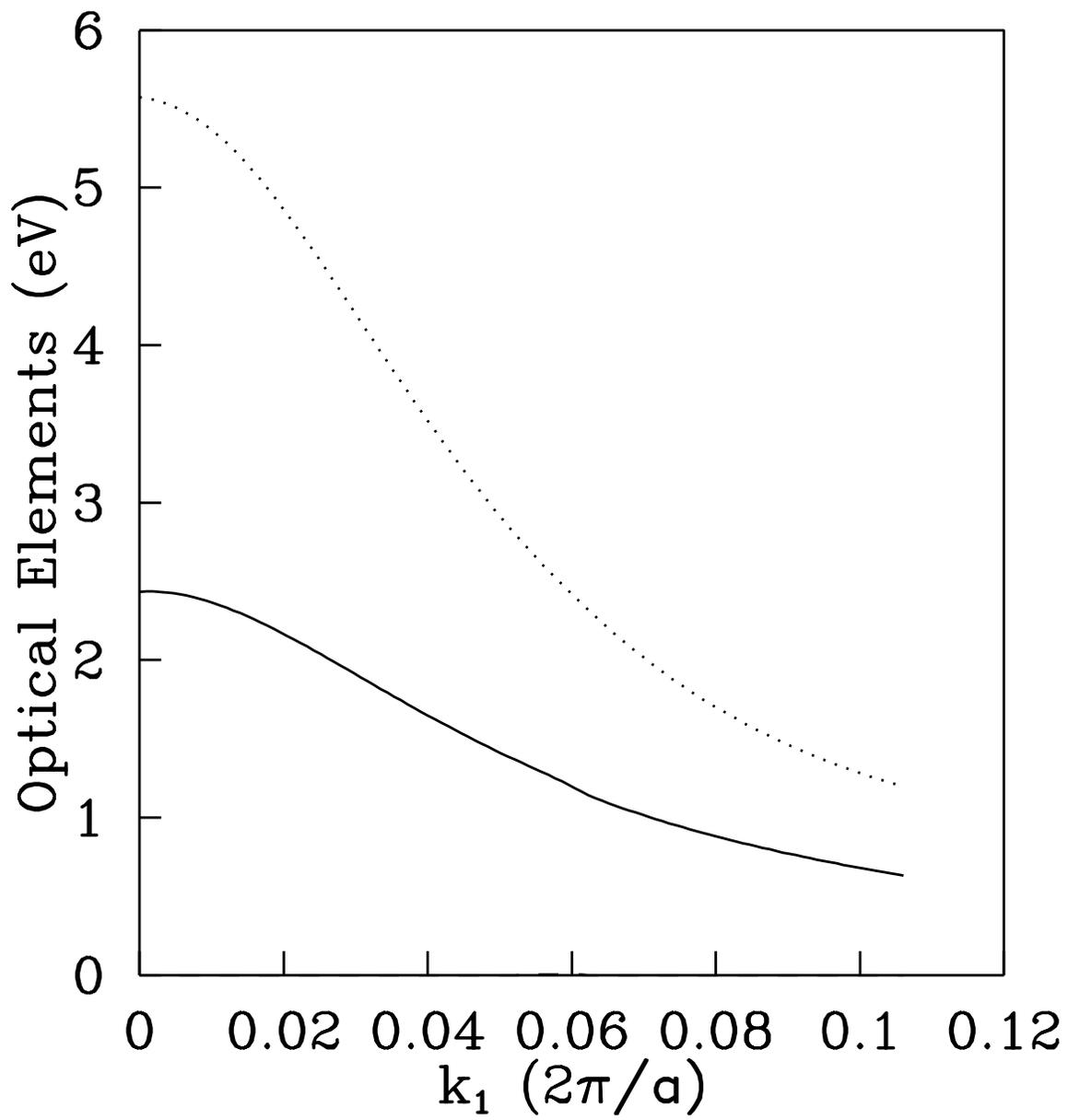
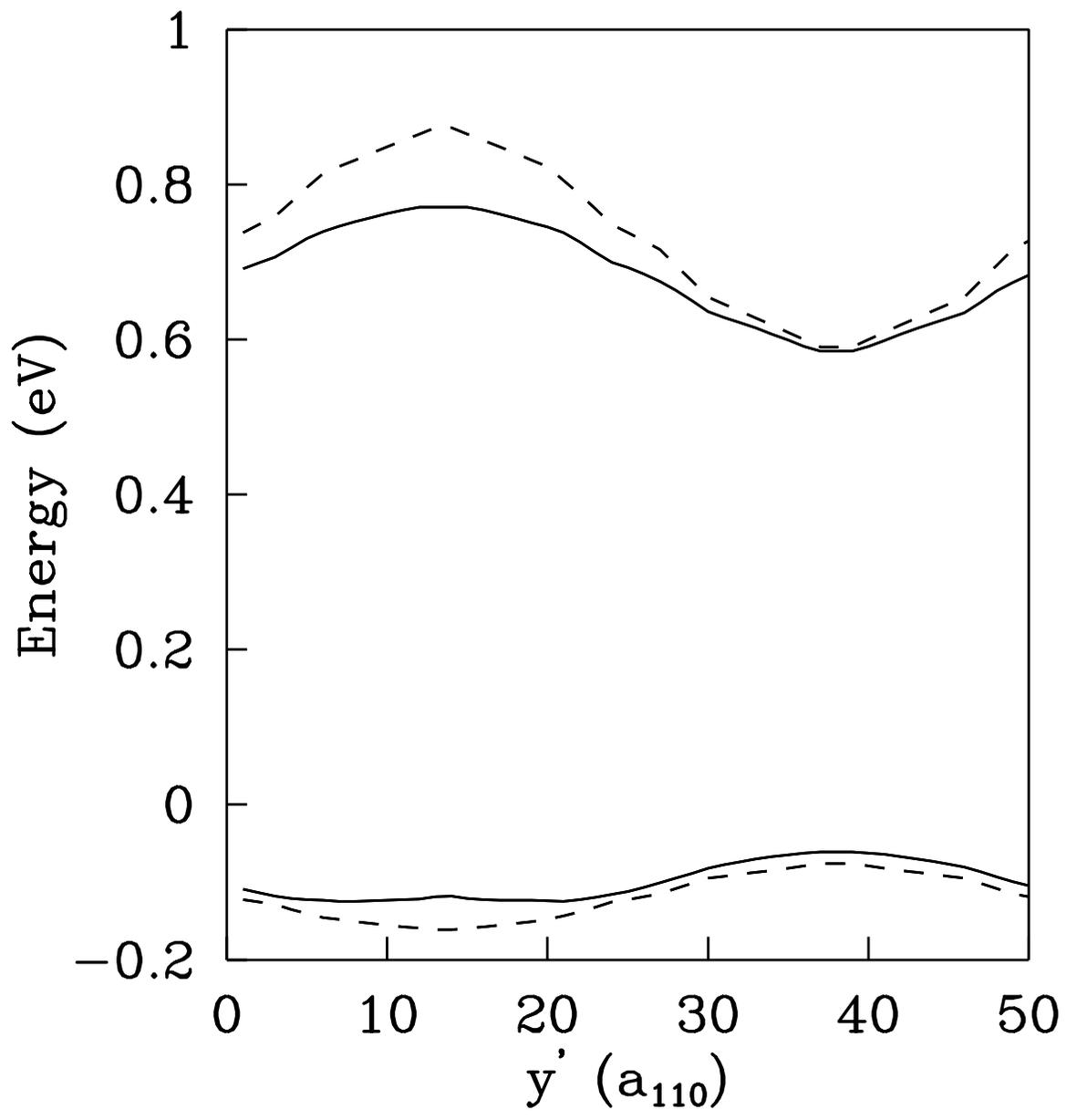
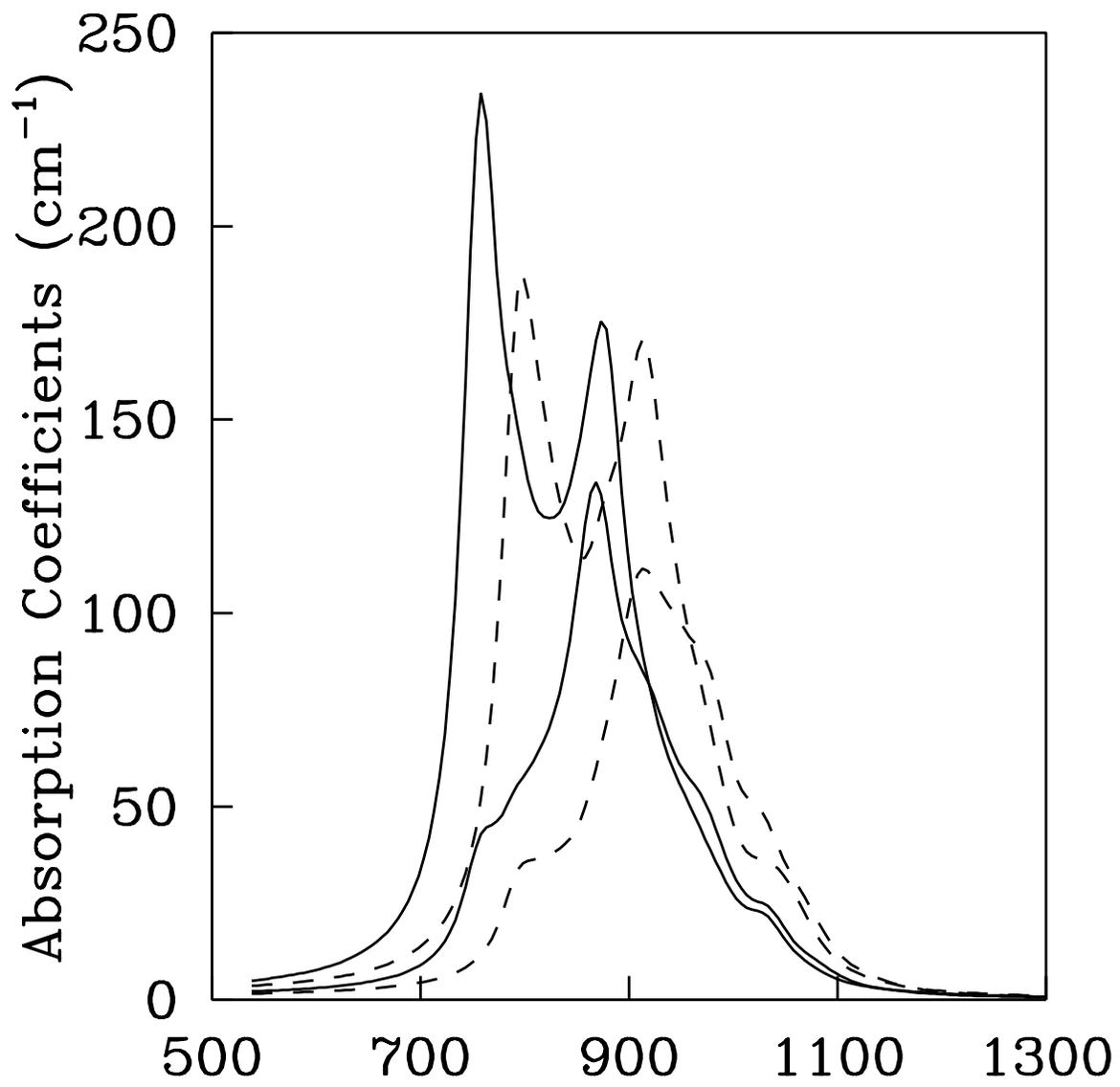
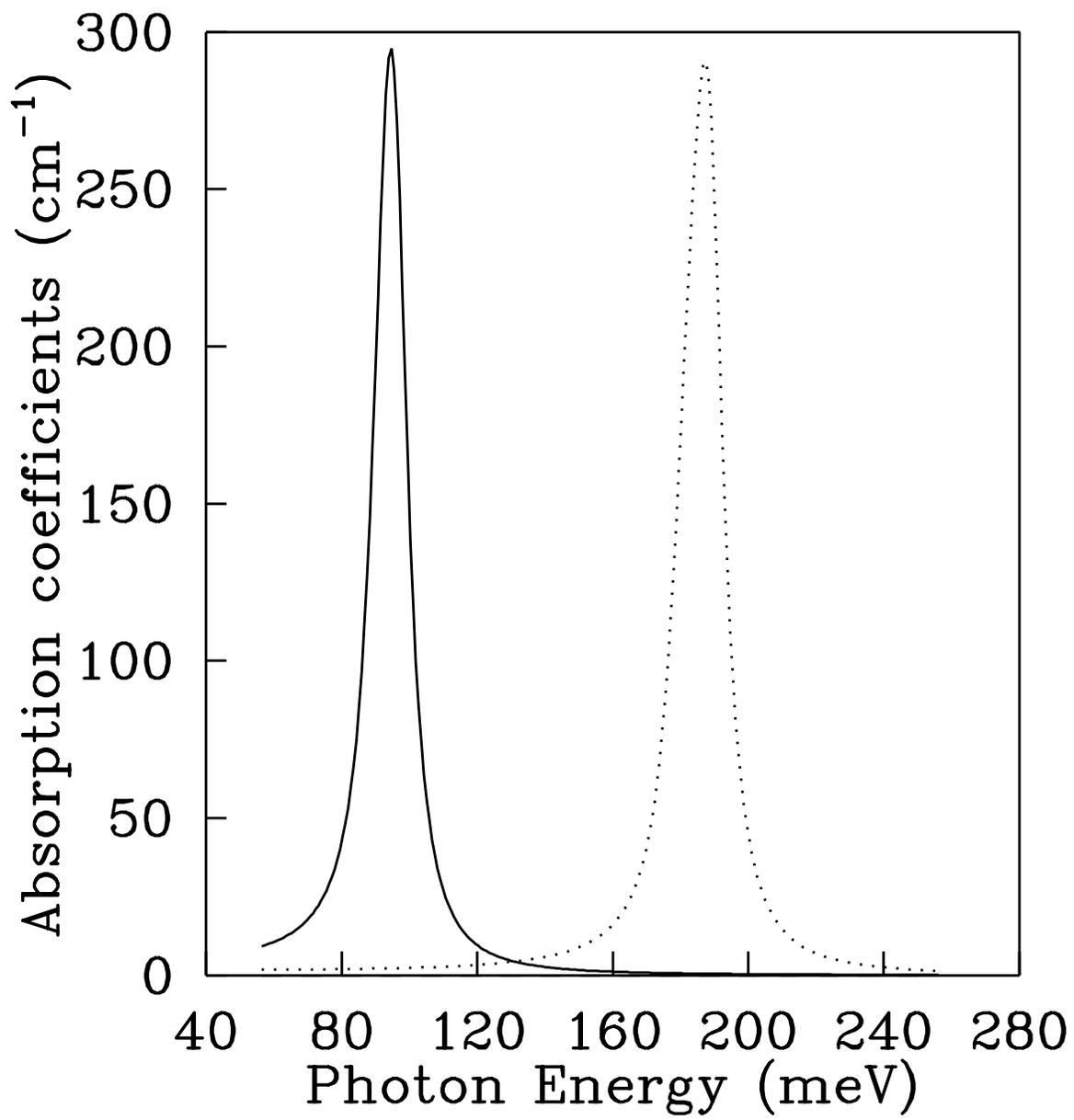
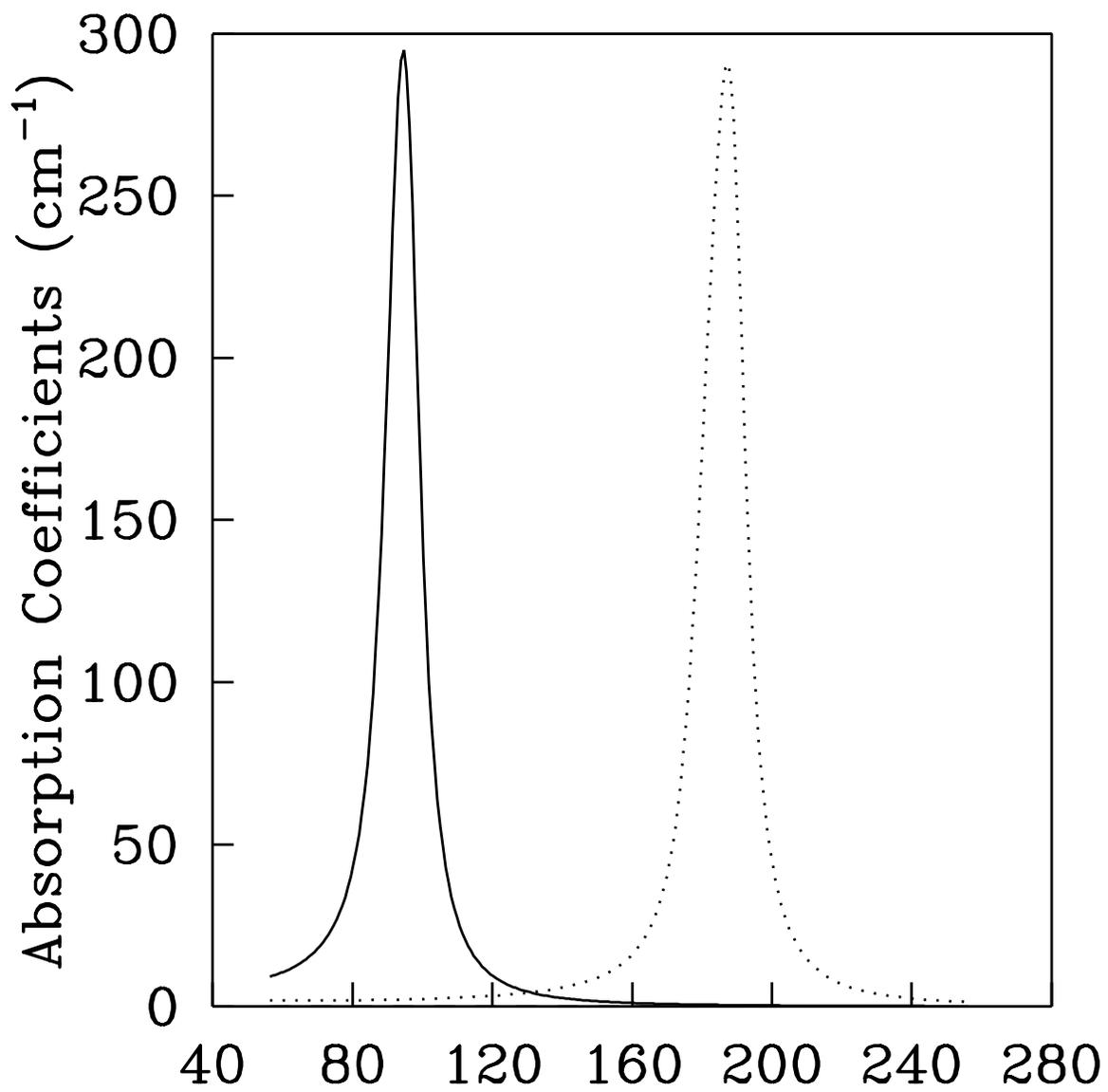
\end{document}